
\input phyzzx
\def\Roman#1{\uppercase\expandafter{\romannumeral #1}.}
\sectionstyle={\Roman}
\def\section#1{\par \ifnum\the\lastpenalty=30000\else
   \penalty-200\vskip\sectionskip \spacecheck\sectionminspace\fi
   \global\advance\sectionnumber by 1
   \xdef\sectionlabel{\the\sectionstyle\the\sectionnumber}
   \wlog{\string\section\space \sectionlabel}
   \TableOfContentEntry s\sectionlabel{#1}
   \noindent {\caps\uppercase\expandafter{\romannumeral\the\sectionnumber
.}\quad #1}\par
   \nobreak\vskip\headskip \penalty 30000 }

\def\subsection#1{\par
   \ifnum\the\lastpenalty=30000\else \penalty-100\smallskip \fi
   \noindent\enspace{#1}\enspace \vadjust{\penalty5000}}

\font\titlefont=cmcsc10 at 14pt
\font\namefont=cmr12
\font\placefont=cmsl12
\font\abstractfont=cmbx12

\font\sxrm=cmr6          \font\sxmi=cmmi6             
  \font\sxsy=cmsy6           
  
\font\nnrm=cmr9          \font\nnmi=cmmi9
  \font\nnsy=cmsy9           \font\nnex=cmex10 at 9pt
  \font\nnit=cmti9           \font\nnsl=cmsl9
  \font\nnbf=cmbx9
\def\nnpt{\def\rm{\fam0\nnrm}%
  \textfont0=\nnrm \scriptfont0=\sxrm \scriptscriptfont0=\fiverm
  \textfont1=\nnmi \scriptfont1=\sxmi \scriptscriptfont1=\fivei
  \textfont2=\nnsy \scriptfont2=\sxsy \scriptscriptfont2=\fivesy
  \textfont3=\nnex \scriptfont3=\nnex \scriptscriptfont3=\nnex
  \textfont\itfam=\nnit \def\it{\fam\itfam\nnit}%
  \textfont\slfam=\nnsl \def\sl{\fam\slfam\nnsl}%
  \textfont\bffam=\nnbf \def\bf{\fam\bffam\nnbf}%
  \normalbaselineskip=11pt
  \setbox\strutbox=\hbox{\vrule height8pt depth3pt width0pt}%
  \let\big=\nnbig \let\Big=\nnBig \let\bigg=\nnbigg \let\Bigg=\nnBigg
  \normalbaselines\rm}
\skip\footins=0.2in
\dimen\footins=4in
\catcode`\@=11
\def\vfootnote#1{\insert\footins\bgroup\nnpt
  \interlinepenalty=\interfootnotelinepenalty
  \splittopskip=\ht\strutbox
  \splitmaxdepth=\dp\strutbox \floatingpenalty=20000
  \leftskip=0pt \rightskip=0pt \spaceskip=0pt \xspaceskip=0pt
  \parfillskip=0pt plus 1fil
  \setbox1=\hbox{*}\parindent=\wd1\let\enspace=\null
  \hangafter1\hangindent\parindent\textindent{#1}\footstrut
  \futurelet\next\fo@t}
\catcode`\@=12
\def\footnoterule{\kern-3pt \hrule width2truein \kern 3.6pt}

\footline={\iftitlepage\hfil\global\titlepagefalse
\else\hss\tenrm\folio\hss\fi}

\newif\iftitlepage

\def\center{\parindent=0pt\leftskip=1in plus 1fill\rightskip=1in plus 1fill}

\def\preprint#1//#2//#3//{\baselineskip=14pt{\rightline{#1}\par}\vskip
.1in{\rightline{#2}\par}
{\rightline{#3}}\medskip}

\def\preprintb#1//#2//{\baselineskip=14pt{\leftline{#1}\par}{\leftline{#2}}}

\def\title#1\par{{\center\baselineskip=16pt
  \titlefont{#1}\par}}
\long\def\author #1// #2// #3// #4//{{\baselineskip=14pt\parskip=0pt
   \center\namefont{#1}\par\placefont#2\par#3\par#4\par}\par}

\long\def\abstract#1//{%
  \centerline{\abstractfont Abstract}\par
  {\baselineskip=14.5pt\advance\leftskip by 0pc\advance\rightskip by
0pc\parindent=10pt
  \def\enspace{\kern.3em}
  \noindent #1\par}}

\def\references{\leftskip=1.2\parindent\parindent=0pt
  \vskip0pt plus.07\vsize\penalty-250\vskip0pt plus-.07\vsize
  \removelastskip\medskip\bigskip\vskip\parskip
  \centerline{REFERENCES}\medskip}

\newcount\refno\refno=0
\long\def\rfrnc#1//{\advance\refno by 1
  $ $\llap{\hbox to\leftskip{\hfil\the\refno.\enspace}}#1\par\smallskip}
\def\PL{{\it Phys.~Lett.}}
\def\NP{{\it Nucl.~Phys.}}
\def\PR{{\it Phys.~Rev.}}

\def\bra#1{\langle #1 \vert}

\def\ket#1{\vert #1 \rangle}

\def\bib#1{\hbox{${}^{#1}$}}

\def\t1{{\tilde 1}}

\def\a3{\alpha_3(m_{Z^0})}

\def\GeV{\,{\rm GeV}}
\def\TeV{\,{\rm TeV}}

\newdimen\jbarht\jbarht=.2pt
\newdimen\vgap\vgap=1pt
\newcount\shiftfactor\shiftfactor=12

\catcode`\@=11
\def\jbarout{\setbox1=\vbox{\offinterlineskip
  \dimen@=\ht0 \multiply\dimen@\shiftfactor \divide\dimen@ 100
    \hsize\wd0 \advance\hsize\dimen@
  \hbox to\hsize{\hfil
    {\multiply\dimen@-2 \advance\dimen@\wd0
    \vrule height\jbarht width\dimen@ depth0pt}%
    \hskip\dimen@}%
  \vskip\vgap\box0\par}\box1}
\def\jbar#1{\mathchoice
  {\setbox0=\hbox{$\displaystyle{#1}$}\jbarout}%
  {\setbox0=\hbox{$\textstyle{#1}$}\jbarout}%
  {\setbox0=\hbox{$\scriptstyle{#1}$}\jbarout}%
  {\setbox0=\hbox{$\scriptscriptstyle{#1}$}\jbarout}}
\catcode`\@=12

\def\simlt{\mathop{\lower.4ex\hbox{$\buildrel<\over\sim$}}}
\def\simgt{\mathop{\lower.4ex\hbox{$\buildrel>\over\sim$}}}

\def\kkb{$K^0-\jbar K^0$}
\def\bbb{$B^0-\jbar B^0$}

\def\ds{$\delta\tilde m^2_{\jbar d_As_B}$}
\def\db{$\delta\tilde m^2_{\jbar d_Ab_B}$}
\def\leaderfill{\leaders\hbox to 1em{\hss\rm .\hss}\hfill}

\advance\vsize by 1truein 
\titlepagetrue

\preprint{\it Submitted to Physics Letters B}//{MIU-THP-92/60}//{December,
1992}//
\vskip 1truein
\title Exact Supersymmetric Amplitude for \kkb\/ and \bbb\/ Mixing

\bigskip
\medskip
\author
John S. Hagelin, S. Kelley and Toshiaki Tanaka//
\vskip.1in
Department of Physics//
Maharishi International University//
Fairfield, IA  52557-1069, USA//

\bigskip
\medskip
\centerline {Abstract}

{\leftskip=.3truein\rightskip=\leftskip\baselineskip=16pt\noindent We present
the most
general supersymmetric amplitude for \kkb\/ and \bbb\/ mixing resulting from
 gluino box diagrams.  We use this amplitude to place general constraints on
the magnitude of
flavor-changing squark mass mixings, and compare these constraints to
theoretical predictions
both in and beyond the Minimal Supersymmetric Standard Model. \par}
\vskip 2.5in
\preprintb{MIU-THP-92/60}//{December, 1992}//

\endpage
\advance\vsize by -1truein 
\baselineskip=18pt
Flavor changing neutral currents (FCNC's) provide an important test of the
radiative structure of the Standard Model and a sensitive probe of new
physics beyond the Standard Model.\bib{1}  Indeed, new
particles proposed by alternative theoretical frameworks can generate
observable
FCNC effects,
even for new particle masses well beyond the range of present
and proposed accelerators.  For many such frameworks, \kkb\/ and \bbb\/ mixing
provide the
most sensitive experimental probe of these FCNC's.\bib{1}

During the past eight years, there have been a number of
theoretical papers exploring FCNC's in supersymmetry.\bib{1-18} These papers
generally
disagree on the magnitude of supersymmetric FCNC's by as much as several
orders of magnitude. These disagreements arise because there has been no
systematic calculation of
the relevant supersymmetric amplitudes for most FCNC processes comparable to
those
that have been performed in the Standard Model.\bib{19} This is because the
supersymmetric contributions are more complicated, involving many new
particles (e.g., six complex charge 2/3 squark fields) which mix in a
complicated way.

In this paper, we present the exact Feynman amplitude for the dominant, gluino
box
contribution to \kkb\/ and \bbb\/ mixing. This calculation may be organized in
either of two
frameworks:\bib{20} first, using exact Feynman rules for mass-eigenstate squark
fields with arbitrary
off-generational squark-gluino couplings; second, as a perturbative expansion
in small,
off-generational squark mass insertions assuming diagonal squark-gluino
couplings (and
non-mass-eigenstate squark fields). The latter formulation affords the most
convenient language both for expressing experimental constraints and for
comparing these
constraints against different theoretical predictions resulting from various
supersymmetric
models.

We have verified that the result in the mass insertion framework can be
systematically derived from
the result in the mass eigenstate basis by expanding around the universal
squark mass.\bib{20}
This provides an important consistency check on the results obtained from both
computational
frameworks.  Since, for $M_{\tilde q}\simlt 1\TeV$, the magnitude of the
expansion parameters,
$\delta\tilde m^2_{AB} / M^2_{\tilde q}$, consistent with experiment are all
much less than one, an
expansion to first order in these small parameters provides an excellent
approximation.  Apart from
algebraic corrections, the main improvements over earlier
computations\bib{11,12} are the
clarification of three distinct terms involving left-right mass insertions
$\Bigl(\bigl[\delta\tilde
m^2_{\jbar d_Ls_R}\bigr]^2$, $\bigl[\delta\tilde m^2_{\jbar d_Rs_L}\bigr]^2$,
$\delta\tilde m^2_{\jbar
d_Ls_R}\delta\tilde m^2_{\jbar d_Rs_L}\Bigr)$ and the identification of how the
experimental
constraints on, and theoretical predictions for, the mass insertions scale with
the supersymmetry
breaking scale.

The dominant supersymmetric contribution to $\Delta S=2$ processes results from
four topologically distinct gluino box diagrams (Figures 1a - d). Neglecting
external
quark momenta compared to heavy internal squark and gluino masses, these
diagrams gives rise to an
effective interaction Lagrangian:
$$\eqalignno{
  {\cal L}^{eff}_{\Delta S=2}&={\alpha^2_s\over 216M^2_{\tilde
q}}\Biggl\{\biggl({\delta\tilde
m_{\jbar d_Ls_L}^2\over M_{\tilde q}^2}\biggr)^2
  [66\tilde f(x)+24xf(x)]
  (\jbar d_i\gamma_{\mu}P_Ls_i)
  (\jbar d_j\gamma^{\mu}P_Ls_j)\cr
&+\biggl( {\delta\tilde m_{\jbar d_Rs_R}^2\over M_{\tilde q}^2}\biggr)^2
  [66\tilde f(x)+24xf(x)]
  (\jbar d_i\gamma_{\mu}P_Rs_i)
  (\jbar d_j\gamma^{\mu}P_Rs_j)\cr
&+\biggl({\delta\tilde m_{\jbar d_Ls_L}^2\over M_{\tilde
q}^2}\biggr)\biggl({\delta\tilde m_{\jbar
d_Rs_R}^2\over M_{\tilde q}^2}\biggr)
  \Bigl([-72\tilde f(x)+504xf(x)](\jbar d_iP_Ls_i)(\jbar d_jP_Rs_j)\cr
 &\hskip1.8in+[120\tilde f(x)+24xf(x)]
  (\jbar d_iP_Ls_j)(\jbar d_jP_Rs_i)\Bigr)\cr
&+\biggl({\delta\tilde m_{\jbar d_Ls_R}^2\over M_{\tilde q}^2}\biggr)^2xf(x)
  \bigl[324(\jbar d_iP_Rs_i)(\jbar d_jP_Rs_j)-108(\jbar d_iP_Rs_j)(\jbar
d_jP_Rs_i)\bigr]\cr
&+\biggl({\delta\tilde m_{\jbar d_Rs_L}^2\over M_{\tilde q}^2}\biggr)^2 xf(x)
  \bigl[324(\jbar d_iP_Ls_i)(\jbar d_jP_Ls_j)-108(\jbar d_iP_Ls_j)(\jbar
d_jP_Ls_i)\bigr]\cr
&+\biggl({\delta\tilde m_{\jbar d_Ls_R}^2\over M_{\tilde
q}^2}\biggr)\biggl({\delta\tilde m_{\jbar
d_Rs_L}^2\over M_{\tilde q}^2}\biggr) \tilde f(x)
  \bigl[108(\jbar d_iP_Ls_i)(\jbar d_jP_Rs_j)\cr
 &\hskip2.1in-324(\jbar d_iP_Ls_j)(\jbar d_jP_Rs_i)\bigr]\Biggr\}&(1)}$$
with
$$\eqalignno{
f(x)&={1\over6(1-x)^5}(-6\ln{x}-18x\ln{x}-x^3+9x^2+9x-17)&{(2a)}\cr
\tilde f(x)&={1\over3(1-x)^5}(-6x^2\ln x-6x\ln{x}+1x^3+9x^2-9x-1)&{(2b)}\cr
\noalign{\smallskip}
x&\equiv M^2_{\tilde g}/M^2_{\tilde q}&{(2c)}\cr}$$
where $M_{\tilde q}$ is the universal (or average) down-squark mass and the
quantity $[66\tilde
f(x)+24xf(x)]\to -1$ for $x=1$.  The soft supersymmetry-breaking FCNC mass
insertions $\delta\tilde
m^2_{\jbar d_As_B}(A,B=L,R)$ appearing in (1) are defined by
$${\cal L}\subset -{\tilde d}^*_A\delta\tilde m^2_{\jbar d_As_B}\tilde
s_B\eqno{(3)}$$
where the squark fields $\tilde d_A$, $\tilde s_B$ reflect a ``super KM"
basis\bib{20} in which the
$\tilde g$, $\tilde\gamma$, $\tilde Z$ gaugino couplings are flavor diagonal
and the
charged-current $\tilde W^\pm$ mixing angles linking quarks and squarks are
equal to the standard KM
angles.

One can easily verify that the interaction Lagrangian (1) reproduces the quark
scattering amplitude
in Figure 1, noting that the following operators give rise to associated
amplitudes:
$$\eqalign{&[\jbar d_\alpha\gamma_\mu P_Ls_\alpha][\jbar d_\beta\gamma^\mu
P_Ls_\beta]\cr
&\to
A=2\bigl(\jbar d_\beta (k_2)\gamma_\mu P_Ls_\alpha (k_1)\bigr)
\bigl(\jbar d_\gamma (k_3)\gamma^\mu P_Ls_\delta (k_4)\bigr)
\bigl(\delta_{\alpha\beta}\delta_{\gamma\delta}+\delta_{\alpha\gamma}\delta_{\beta\delta}
\bigr)\cr}
\eqno(4a)$$
$$\eqalign{[\jbar d_iP_As_j][\jbar d_kP_Bs_l]
&\to
A=\bigl(\jbar d_\beta (k_2)P_As_\alpha (k_1)\bigr)\bigl(\jbar d_\gamma
(k_3)P_Bs_\delta (k_4)\bigr)
\delta_{j\alpha}\delta_{i\beta}\delta_{k\gamma}\delta_{l\delta}\cr
&-\bigl(\jbar d_\gamma (k_3)P_As_\alpha (k_1)\bigr)\bigl(\jbar d_\beta
(k_2)P_Bs_\delta
(k_4)\bigr) \delta_{j\alpha}\delta_{i\gamma}\delta_{k\beta}\delta_{l\delta}\cr
&-\bigl(\jbar d_\beta (k_2)P_As_\delta (k_4)\bigr)\bigl(\jbar d_\gamma
(k_3)P_Bs_\alpha
(k_1)\bigr) \delta_{j\delta}\delta_{i\beta}\delta_{k\gamma}\delta_{l\alpha}\cr
&+\bigl(\jbar d_\gamma (k_3)P_As_\delta (k_4)\bigr)\bigl(\jbar d_\beta
(k_2)P_Bs_\alpha
(k_1)\bigr)
\delta_{j\delta}\delta_{i\gamma}\delta_{k\beta}\delta_{l\alpha}\cr}\eqno(4b)$$
The form of ${\cal L}^{eff}_{\Delta S=2}$ in the ``minimal susy limit" where
$\delta\tilde m^2_{LL}$
dominates differs from previous analyses,\bib{10,11} which reported color octet
operator structures in
addition to the color singlet $(ii)(jj)$ operator shown in (1).  (A color octet
left-left operator
can always be converted to a color singlet operator by a Fierz transformation.)
 One can use the
effective Lagrangian (1) to assess  the effects of off-generational mass
mixings \ds of  arbitrary
chirality which are radiatively induced in minimal supersymmetry or which may
appear  in non-minimal
supersymmetric models. The $\Delta B=2$ effective interaction Lagrangian can be
read directly from
(1) by the substitution \ds $\to$ \db.

The application of the low-energy effective $\Delta S=2$ ($\Delta B=2$)
interaction Lagrangians derived above to \kkb\/ (\bbb) mixing requires
estimating the
matrix elements of the various operators in ${\cal L}^{eff}$ between initial
and final
state mesons.  The estimation of such hadronic matrix elements is notoriously
difficult,
and is generally accompanied by large uncertainties due to long-distance,
non-perturbative strong-interaction physics.  There are two factors which
mitigate the effect of
these hadronic uncertainties in the current phenomenological context:

1) The dominant supersymmetric contribution to \kkb\/ and \bbb\/ mixing in the
Minimal
Supersymmetric Standard Model (MSSM)\bib{21} results from a $\delta\tilde
m^2_{LL}$ mass insertion,
which gives rise to the same $(V-A)^2$ operator structure found in the Standard
Model.  This makes
comparison of supersymmetric and Standard Model contributions relatively
straightforward;

2) For the \bbb\/ system, the valence quark approximation employed below is
expected to be a good approximation.
This belief is supported by lattice monte-carlo estimates of the \bbb\/
matrix element ``fudge factor" $B_B$ which give $B_B\simeq1$.\bib{22}

The conventional result for the standard $(V-A)^2$ $\Delta
S=2$ operator is
$$\bra{K^0}[\jbar d_i\gamma_{\mu}P_Ls_i][\jbar
d_j\gamma^{\mu}P_Ls_j]\ket{\jbar
K^0}={2\over3}f_K^2m_K^2B_K^a\eqno(5a)$$
where $f_K\simeq 165$ MeV is the K-meson decay constant and $B^a_K=1$
corresponds to the
``vacuum insertion" result.  Various estimates of this matrix element place
$B_K^a$ in
the range of $0.3-1.0$,\bib{23} with a value $B_K^a\sim0.7$ favored by lattice
gauge
results.\bib{22}  Matrix elements for the other hadronic operators appearing in
${\cal
L}^{eff}_{\Delta S=2}$ follow from current algebra:
$$\eqalignno{
\bra{K^0}[\jbar d_\alpha P_Ls_\alpha ][\jbar d_\beta P_Ls_\beta]\ket{\jbar K^0}
  &={5\over12}\left( {m_K\over m_s+m_d}\right)^2f_K^2m_K^2&{(5b)}\cr
 \noalign{\hbox{}}
\bra{K^0}[\jbar d_\alpha P_Ls_\beta][\jbar d_\beta P_Ls_\alpha]\ket{\jbar K^0}
  &=-{1\over{12}}\left({m_K\over m_s+m_d}\right)^2f_K^2m_K^2&{(5c)}\cr
 \noalign{\hbox{}}
\bra{K^0}[\jbar d_\alpha P_Ls_\alpha][\jbar d_\beta P_Rs_\beta]\ket{\jbar K^0}
  &=\left\{{1\over12}-{1\over2}
   \left({m_K\over m_s+m_d}\right)^2\right\}f_K^2m_K^2&{(5d)}\cr
 \noalign{\hbox{}}
\bra{K^0}[\jbar d_\alpha P_Ls_\beta][\jbar d_\beta P_Rs_\alpha]\ket{\jbar K^0}
  &=\left\{ {1\over4}-{1\over6}\left( {m_K\over m_s+m_d}\right)^2 \right\}
    f_K^2m_K^2&{(5e)}\cr}$$
Similar fudge factors $B_K^{b-e}$ should be associated
with each of the matrix elements (5b-e) above.  Note that the matrix elements
(5b-e) are enhanced relative to (5a) by a factor $\bigl({m_K\over
m_s+m_d}\bigr)^2\sim10$. This enhancement, which is significant for (5b and d),
combined with large numerical coefficients for these non-standard chiral
contributions
to ${\cal L}^{eff}_{\Delta S=2}$ (1), raise the prospect of interesting
enhancements to $\Delta M_K$ outside of the MSSM, for which we will find that
the
$\delta\tilde m^2_{LL}$ mass insertion gives the dominant contribution.  The
corresponding results for
the $\Delta B=2$ matrix elements follow from (5) by setting $B_B\simeq
\bigl({m_B\over
m_b+m_d}\bigr)^2\simeq1$.
\vskip.2in
\vbox{\tabskip=0pt \offinterlineskip
\def\tablerule{\noalign{\hrule}}
\def\struta{\vrule height12pt depth 5pt width 0pt}
\def\strutb{\vrule height15pt depth 10pt width 0pt}
\def\strutc{\vrule height19pt depth 10pt width 0pt}

\halign to 418pt{\struta#& \vrule#\tabskip=1em plus2em&
  \hfil#\hfil& \vrule#&\hfil#\hfil& \vrule#&\hfil#\hfil&
  \vrule#&\hfil#\hfil& \vrule#\tabskip=0pt\cr\tablerule
\strutb&&\multispan7\hfil Phenomenological Upper Bounds on \ds and \db
\hfil&\cr\tablerule
\strutc&&\omit&&\multispan3\hfil
   $\Bigl({\delta\tilde m^2_{\jbar d_As_B}\over M^2_{\tilde q}}\Bigr)
   \Bigl({\delta\tilde m^2_{\jbar d_{A'}s_{B'}}\over M^2_{\tilde q}}\Bigr)$
   \hfil&&
   $\Bigl({\delta\tilde m^2_{\jbar d_Ab_B}\over M^2_{\tilde q}}\Bigr)
   \Bigl({\delta\tilde m^2_{\jbar d_{A'}b_{B'}}\over M^2_{\tilde q}}\Bigr)$
   \hfil&\cr\tablerule
&&$(\alpha\beta)(\alpha'\beta')$&&$\Delta M_K$&&$\epsilon_K$&&$\Delta M_B$&\cr
\tablerule
&&$(LL)^2$ or $(RR)^2$&&$(0.10)^2$&&$(0.0080)^2$&&$(0.27)^2$&\cr\tablerule
&&$(LL)(RR)$&&$(0.0060)^2$&&$(0.00049)^2$&&$(0.073)^2$&\cr\tablerule
&&$(LR)^2$ or $(RL)^2$&&$(0.0082)^2$&&$(0.00066)^2$&&$(0.082)^2$&\cr\tablerule
&&$(LR)(RL)$&&$(0.044)^2$&&$(0.0035)^2$&&$(0.14)^2$&\cr\tablerule}
{\leftskip=.2truein\rightskip=\leftskip\tenpoint\noindent TABLE I. All numbers
in
the table must be multiplied by the factor $(M_{\tilde q}/{\rm TeV})^2$.
Numerical values assume
$\sqrt{x}\sim M_{\tilde g}/M_{\tilde q}=1$.  Stricter (weaker) results
generally apply to
$x<1$ ($x>1$).  Bounds derived from $\epsilon_K$ assume maximal CP violation.
Bounds from $\Delta
M_B$ must be scaled by (160 MeV/$f_B$).\par}}

Phenomenological constraints on the various FCNC mass mixings \ds \break
appearing in ${\cal
L}^{eff}_{\Delta S=2}$ can be derived by inserting the matrix elements (5) into
${\cal
L}^{eff}_{\Delta S=2}$ and requiring $\Delta M_K^{\rm susy}\le \Delta
M_K^{exp}$ for each of the
chiral contributions to the $K_L-K_S$ mass difference: $$\Delta M_K\simeq
2M_{12}={1\over
m_K}\bra{K^0}-{\cal L}^{eff}_{\Delta S=2}\ket{\jbar K^0}=3.52\times
10^{-15}\GeV\eqno(6)$$
Barring accidental cancellations between the various chiral
contributions, this constraint is conservative in that the short-distance,
Standard Model charm-quark contribution is itself of order $\Delta M_K^{exp}$,
and of the
correct sign.  Setting $B_K^a=0.7$ (only) and taking $x=1$
yields the constraints on \ds due to $\Delta M_K$ shown
in Table I.

The results of Table I are significant and warrant comment. All the
constraints from $\Delta M_K$ are useful constraints (i.e. $\delta \tilde
m_{AB}^2/M^2_{\tilde q}<1$), even for heavy squark masses $M_{\tilde q}\gg1$
TeV. Indeed the
constraint on $\delta \tilde m_{LL}^2$, which is actually the weakest of the
constraints from $\Delta M_K$, provides useful information on the down-squark
mass
matrix for $M_{\tilde q, \tilde g}$ as large as 10 TeV. Above 10 TeV, the upper
bound on $\delta
\tilde m_{\bar d_Ls_L}^2/M_{\tilde q}^2$ rises above unity, at which point the
perturbative expansion
in powers of $\delta \tilde m^2/M_{\tilde q}^2$ breaks down, and no further
useful information
remains on the form of the squark mass matrix.

The constraint for $\delta \tilde m_{RR}^2$ is identical to that for $\delta
\tilde
m_{LL}^2$. However, Table I shows that in the presence of {\it both} types of
mass
insertions, even stronger constraints apply. This is because the numerical
coefficients
(e.g., $504$) in ${\cal L}^{eff}_{\Delta S=2}$ (1) and the operator matrix
element (5)
$\propto\big[ m_K/(m_s+m_d)\big]^2$ are both enhanced relative to the $\delta
\tilde m_{LL,RR}^2$ contributions. This results in useful constraints (i.e.
$\delta \tilde
m_{LL,RR}^2/M_{\tilde q}^2<1$) on the squark mass
matrix for sparticle masses $M_{\tilde q, \tilde g}$ greater than 100 TeV. For
the same
reasons, the bounds on $\delta \tilde m_{\bar d_Ls_R}^2$ and $\delta \tilde
m_{\bar
d_Rs_L}^2$ are similarly stringent and furnish useful constraints for sparticle
masses
$M_{\tilde q, \tilde g}$ up to 100 TeV.

The most stringent bounds on \ds follow from the CP
impurity parameter $\epsilon_K$.  Table I shows the
constraints on the \ds which result from
setting $\epsilon_K^{susy}<\epsilon_K^{exp}$ and assuming maximal CP-violating
phases for each of
the chiral contributions. Note that, in many cases, CP violation in \kkb\/ is
sensitive to
supersymmetric contributions from sparticle masses $M_{\tilde q,\tilde g}>1000$
TeV.

The analysis of $\Delta M_B$ is identical to $\Delta M_K$, with appropriate
substitution of flavor indices.  Although numerically less stringent than those
from $\Delta M_K$,
the constraints in Table I can be equally restrictive for many supersymmetric
models (like the MSSM)
which predict \db $\gg$ \ds.\bib{20}

In the MSSM with universal soft supersymmetry breaking, FCNC mass insertions
are generated by
renormalization effects between the unification scale $M_U$ and $m_W$.\bib{2-5}
 The
resulting low-energy mass insertions are predominantly left-left, and exhibit
the following
approximate flavor dependence:\bib{20}
$${[\delta\tilde m^2_{\jbar d_Ld_L}]_{ij}\over M_{\tilde q}^2}
  =c_{LL}(\xi_0,\xi_A)[V^\dagger\lambda_u^2 V]_{ij}\eqno(7)$$
where $\lambda_u$ is the diagonalized charge 2/3 quark mass matrix and $V$ is
the KM matrix.  The
renormalization-group coefficient $c_{LL}$ is flavor independent, and is
plotted in Figure 2 as a
function of $\xi_0\equiv m_0/m_{1/2}$ and $\xi_A\equiv A/m_{1/2}$, where $m_0$,
$m_{1/2}$, $A$ are
respectively the universal soft supersymmetry-breaking scalar mass, gaugino
mass and trilinear
coupling at $\mu=M_U$.  Using the experimental upper bound\bib{24} $V^*_{td}
V_{ts}<7.3\times
10^{-4}$ on the KM angles associated with the dominant $\lambda^2_t$
contribution, we observe that
the radiatively-generated $\delta\tilde m^2_{\jbar d_Ls_L}$ is too small to
contribute significantly
to $\Delta M_K$ or $\epsilon_K$.

Such small FCNC results are a relatively unique prediction of the MSSM.
Most nontrivial extensions of the MSSM contain extra Yukawa couplings, and
these
generically lead to large FCNC's.\bib{6}  In supersymmetric Flipped
SU(5),\bib{25} for example, we
find potentially large FCNC's generated above the GUT scale:\bib{20}
$$\eqalignno{
[\delta\tilde m_{\jbar d_L d_L}^2]_{ij}
&=-{1\over 8\pi^2}[\lambda_6^*\lambda_6^T]_{ij}
\ln\biggl({M_{Pl}\over M_{GUT}}\biggr)[3m^2_0+A^2]&(8a)\cr
\noalign{\hbox{}}
[\delta\tilde m_{\jbar d_R d_R}^2]_{ij}
&=[\delta\tilde m_{\jbar d_L d_L}^2]_{ij}^*&(8b)\cr
\noalign{\hbox{}}
[\delta\tilde m_{\jbar d_R d_L}^2]_{ij}
&=-\eta^{1}_{ij}v={vA\over 8\pi^2}
[\lambda_6\lambda_6^\dagger\lambda_d
+\lambda_d\lambda_6^*\lambda_6^T]_{ij}\ln\biggl({M_{Pl}\over
M_{GUT}}\biggr)&(8c)\cr
\noalign{\hbox{}}
[\delta\tilde m_{\jbar d_L d_R}^2]_{ij}
&=[\delta\tilde m_{\jbar d_R d_L}^2]_{ij}^*
&(8d)\cr
}$$
where $\lambda_6$ is an {\it a-priori} unknown Yukawa coupling associated with
a see-saw neutrino
mass mechanism.\bib{25}

 From the phenomenological constraints in Table I, we can derive useful\break
bounds on
the unknown GUT Yukawa $\lambda_6$.  The strongest bounds come from the
$(\delta\tilde
m^2_{LL})(\delta\tilde m^2_{RR})$ contributions to $\Delta M_K$, $\Delta M_B$,
from
which we obtain: $$\eqalignno{
\bigl|[\lambda_6^*\lambda_6^T]_{12}\bigr|\ln{\Bigl({M_{Pl}\over M_{GUT}}\Bigr)}
&<0.47{\xi_0^2+6\over 3\xi_0^2+\xi_A^2}\Bigl({M_{\tilde q}\over 1\TeV}\Bigr)
&(9a)\cr
\noalign{\hbox{}}
\bigl|[\lambda_6^*\lambda_6^T]_{13}\bigr|\ln{\Bigl({M_{Pl}\over M_{GUT}}\Bigr)}
&<1.28{\xi_0^2+6\over 3\xi_0^2+\xi_A^2}\Bigl({M_{\tilde q}\over 1\TeV}\Bigr)
&(9b)\cr
}$$
where we have used the approximate low-energy relation $M^2_{\tilde q}\simeq
(\xi^2_0+6)
m^2_{1/2}$.\bib{20}

Equation (9) provides important knowledge
about the unknown matrix $\lambda_6$ and the pattern of soft supersymmetry
breaking.
For instance, if supersymmetry breaking takes the form of either $m_0$ or A,
then
(9) requires certain elements of $\lambda_6$ to be quite small---smaller
than expected from superstring theories, which relate Yukawa couplings to
gauge couplings $\lambda\approx g\approx 0.7$ at $\mu=M_{Pl}$.
One may be forced to conclude that these couplings vanish at the tree level
in such theories---or that soft
supersymmetry breaking takes the form of $m_{1/2}$.

Note that because of (8b), the Flipped contribution to CP violation
$\epsilon_K$ vanishes.  Flipped
SU(5) can, however, contribute significantly to ``direct" CP violation
$\epsilon'$, $K_L\to\pi^0ee$
and $K_L\to\pi^0\nu\jbar\nu$, as shown in ref. 20.

We conclude that FCNC processes, and \kkb\ and \bbb\ mixing in particular,
provide a sensitive probe of physics beyond
the MSSM.  In the presence of
sparticles $M_{\tilde q}, M_{\tilde g}\lsim 1000\TeV$, these FCNC processes
provide an
experimental window on the nature of physics at extremely high energies which
should clearly be
exploited to the fullest extent possible.

The work of one of us (J.S.H.) was supported in part by the National Science
Foundation under grant no. PHY-9118320.

\references
\raggedright
\tenpoint
\advance\baselineskip by -3pt
\rfrnc
For a review, see J.S.~Hagelin and L.S.~Littenberg, {\it
Prog.~Part.~Nucl.~Phys.~}{\bf 23}
(1989) 1.//

\rfrnc
J.F.~Donoghue, H.P.~Nilles and D.~Wyler, \PL\ {\bf 128B} (1983) 55.//

\rfrnc
F.~Borzumati and A.~Masiero, {\it Phys.~Rev.~Lett.~}{\bf 57}
(1986) 961.//

\rfrnc
A.~Bouquet, J.~Kaplan and C.A.~Savoy, \PL\ {\bf 148B} (1984) 69.//

\rfrnc
M.J.~Duncan, \NP\ {\bf B221} (1983) 285.//

\rfrnc
L.J.~Hall, V.A.~Kostelecky and S.~Raby, \NP\ {\bf B267} (1986) 415.//

\rfrnc
J.-M.~Gerard, W.~Grimus, A.~Masiero, D.V.~Nanopoulos and A.~Raychaudhuri, {\it
Nucl.
Phys.}\ {\bf B253} (1985) 93.//

\rfrnc
M.~Dugan, B.~Grinstein and L.~Hall, \NP\ {\bf B255} (1985) 413.//

\rfrnc
J.-M.~Gerard, W.~Grimus, A.~Masiero, D.V.~Nanopoulos and
A.~Raychauduri, {\it Phys. Lett.}\ {\bf 141B} (1984) 79;\break
P.~Langacker and R.~Satiapalan, \PL\ {\bf 144B} (1984) 401.//

\rfrnc
S.~Bertolini, F.~Borzumati and A.~Masiero, \PL\ {\bf 194B} (1987) 545 and
Erratum, and
{\bf 198B} (1987) 590.//

\rfrnc
L.I.~Bigi and F.~Gabbiani, \NP\ {\bf B352} (1991) 309.//

\rfrnc
F.~Gabbiani and A.~Masiero, \NP\ {\bf B322} (1989) 235.//

\rfrnc
J.-M.~Gerard, W.~Grimus, A.~Raychaudhuri and G.~Zoupanos, \PL\ {\bf 140B}
(1984) 349.//

\rfrnc
J.-M.~Gerard, W.~Grimus and A.~Raychaudhuri, \PL\ {\bf 145B} (1984) 400.//

\rfrnc
S.~Bertolini, F.~Borzumati and A.~Masiero, \PL\ {\bf 192B} (1987) 437;\break
T.M.~Aliev and M.I.~Dobroliubov, \PL\ {\bf 237B} (1990) 573.//

\rfrnc
T.~Inami and C.S.~Lim, \NP\ {\bf B207} (1982) 533;\break
J.-M.~Frere and M.B.~Gavela, \PL\ {\bf 132B} (1983) 107;\break
G.~Altarelli and P.J.~Franzini, {\it Z.~Phys.~}{\bf C37} (1988) 271;\break
E.~Franco and M.~Mangano, \PL\ {\bf 135B} (1984) 445.//

\rfrnc
M.J.~Duncan and J.~Trampetic, \PL\ {\bf 134B} (1984) 439.//

\rfrnc
B.A.~Campbell, \PR\ {\bf D28} (1983) 209;\break
F.~Del~Aguila, J.A.~Grifols, A.~Mendez, D.V.~Nanopoulos and M.~Srednicki, \PL\
{\bf 129B}
(1983) 77.//

\rfrnc
T.~Inami and C.S.~Lim, {\it Prog.~Theor.~Phys.~}{\bf 65} (1981) 297.//

\rfrnc
J.S.~Hagelin, S.~Kelley and T.~Tanaka, Maharishi International University
preprint MIU-THP-92/59
(1992), submitted to {\it Nucl. Phys. B}.//

\rfrnc
For a review of the MSSM see, e.g., L.~Ibanez and G.G.~Ross, {preprint
CERN-TH 6142/92 and references therein.}//

\rfrnc
M.B.~Gavela, L.~Maiani, S.~Petrarca and F.~Rapuano, \NP\ {\bf B306} (1988)
677;\break
C.R.~Allton, C.T.~Sachrajda, V.~Lubicz, L.~Maiani and G.~Martinelli, \NP\ {\bf
B349}
(1991) 598.//

\rfrnc
J.F.~Donoghue, E.~Golowich and B.R.~Holstein, \PL\ {\bf 119B} (1982) 412;\break
A.~Pich and E.~de Rafael, \PL\ {\bf 158B} (1985) 477;\break
R.~Decker, \NP\ {\bf B277} (1986) 661;\break
N.~Bilic, C.A.~Dominguez and B.~Guberina, {\it Z. Phys.} {\bf C39} (1988)
351;\break
W.A.~Bardeen, A.J.~Buras and J.-M.~Gerard, \PL\ {\bf 211B} (1988) 343.//

\rfrnc
Y.~Nir, preprint SLAC-PUB 5874 (1992).//

\rfrnc
I.~Antoniadis, J.~Ellis, J.S.~Hagelin and D.V.~Nanopoulos, \PL
{\bf 194B} (1987) 231.//\par

\end